\documentclass[reprint,superscriptaddress,pra]{revtex4-1}
\pdfoutput=1

\usepackage{graphicx}
\usepackage[space]{grffile}
\usepackage{amsmath,amssymb,amstext}
\usepackage{xcolor}

\usepackage{bibunits}

\usepackage{ifpdf}
\ifpdf
	\usepackage{epstopdf}
	\pdfpageattr {/Group << /S /Transparency /I true /CS /DeviceRGB>>}
\fi

\newcommand{\Rb}{${}^{87}\textrm{Rb}$ }

\begin{document}
\title{Thermometry and cooling of a Bose-Einstein condensate to 0.02 times the critical temperature}
\author{Ryan Olf}
\email{ryan@efrus.com}
\affiliation{Department of Physics, University of California, Berkeley, California 94720, USA}
\author{Fang Fang}
\affiliation{Department of Physics, University of California, Berkeley, California 94720, USA}
\author{G. Edward Marti}
\altaffiliation[Present address: ]{JILA, National Institute of Standards and Technology and University of Colorado, Boulder, Colorado 80309-0440, USA; edward.marti@jila.colorado.edu}
\affiliation{Department of Physics, University of California, Berkeley, California 94720, USA}
\author{Andrew MacRae}
\affiliation{Department of Physics, University of California, Berkeley, California 94720, USA}
\author{Dan M. Stamper-Kurn}
\affiliation{Department of Physics, University of California, Berkeley, California 94720, USA}
\affiliation{Materials Sciences Division, Lawrence Berkeley National Laboratory, Berkeley, California 94720, USA}
\date{May 21, 2015} 

\maketitle

\textbf{Ultracold gases promise access to many-body quantum phenomena at convenient length and time scales. However, it is unclear whether the entropy of these gases is low enough to realize many phenomena relevant to condensed matter physics, such as quantum magnetism.  Here we report reliable single-shot temperature measurements of a degenerate \Rb gas by imaging the momentum distribution of thermalized magnons, which are spin excitations of the atomic gas.  We record average temperatures as low as $0.022(1)_\mathrm{stat}(2)_\mathrm{sys}$ times the Bose-Einstein condensation temperature, indicating an entropy per particle, $S/N\approx0.001\, k_B$ at equilibrium, that is well below the critical entropy for antiferromagnetic ordering of a Bose-Hubbard system.  The magnons themselves can reduce the temperature of the system by absorbing energy during thermalization and by enhancing evaporative cooling, allowing low-entropy gases to be produced within deep traps.}

\begin{bibunit}[naturemagnourl]

Trapped quantum gases can be brought to impressively low temperatures.  A single-component Bose gas was cooled to around 500 pK by adiabatic expansion \cite{lean03pico}, and a two-component lattice-trapped Bose gas was cooled to 350 pK using a spin demagnetization technique \cite{medl11demag}.  However, the entropies achieved in these systems are still much higher than would be required to observe many-body quantum effects such as magnetic ordering or $d$-wave superconductivity of atoms in optical lattices \cite{bloc12qsim}.  Estimated critical entropies required to achieve quantum magnetic ordering of atoms in optical lattices include $S/N \sim 0.3 \, k_B$ and $S/N \sim 0.03 \, k_B$ for bosons in state-dependent three- and two-dimensional lattices, respectively \cite{capo10critical}, and similar values for N\'{e}el ordering of lattice-trapped Fermi gases \cite{Paiva2011}, with lower entropy needed for less strongly interacting gases.

In many experiments on strongly interacting atomic-gas systems, the low-entropy regime is reached by first preparing a weakly interacting bulk Bose gas at the lowest possible temperature, and then slowly transforming the system to become strongly interacting \cite{Greiner2002,Bakr2010,Paredes2004,Trotzky2008,Spielman2007}. To discern whether the transformation is adiabatic and to determine indirectly the thermodynamic properties of the strongly interacting system, the system is returned to the weakly interacting regime where relations between temperature, entropy, and other properties are known. Therefore, methods to lower entropies and measure temperatures of weakly interacting gases are important for the study of both weakly and strongly interacting atomic-gas systems. 

In this Letter, we report cooling a Bose gas to a few percent of the condensation temperature, $T_c$, corresponding to an entropy per particle of $10^{-3}\,k_B$ ($k_B$ is the Boltzmann constant), nearly two orders of magnitude lower than the previous best in a dilute atomic gas \cite{ku12lambda,medl11demag}. Surprisingly, we achieve this low entropy using a standard technique: forced evaporation in an optical dipole trap, which we find remains effective in a previously uncharacterized regime. The lowest temperatures we report are achieved at very shallow final trap depths, as low as 20 nK, set by stabilizing the optical intensity with a long-term fractional reproducibility better than $10^{-2}$. In addition, we demonstrate and characterize a method of cooling that lowers the entropy without changing the trap depth, possibly allowing the low entropy regime to be reached or maintained in systems where the trap depth is constrained.

Both thermometry and cooling require a means of distinguishing thermal excitations.  For example, forced evaporative cooling \cite{luit96,kett96evap} depends on the ability to selectively expel high-energy  excitations from the system.  Similarly, thermometry of a degenerate quantum gas requires one to identify the excitations that distinguish a zero- from a non-zero-temperature gas. Both these tasks become difficult when $S/N$, or, equivalently, the fraction of thermal excitations $N_\mathrm{th}/N$, is small \cite{McKay2011}.  Time-of-flight temperature measurements, in which the gas is released from the trap and allowed to expand before being imaged, have required $N_\mathrm{th}/N$ of at least several percent, limiting such thermometry of Bose gases to $T \geq 0.3 \, T_c$ and of Fermi gases to $T \geq 0.05 \, T_F$, where $T_c$ and $T_F$ are the Bose-Einstein    
condensation and Fermi temperatures, respectively.

We extend thermometry to the deeply degenerate regime of a Bose gas by measuring the momentum distribution of a small number of spin excitations, similar to the co-trapped impurity thermometry commonly employed in Fermi gases \cite{Regal2007,Nascimbene2010,Spiegelhalder2009,McKay2010}. Even in a highly degenerate Bose gas with a vanishingly small non-condensed fraction, the minority spin population can be made dilute enough to remain non-degenerate and thereby carry a large entropy and energy per particle. Furthermore, the minority spins that we use---magnons within a ferromagnetic spinor Bose-Einstein condensate---support a higher $N_\mathrm{th}$ than the majority spins because of their free-particle density of states \cite{Marti2014}, increasing the signal of the temperature measurement. Performing spin-selective measurements on the minority spins allows the gas temperature to be readily determined.

\begin{figure}[tb!]
\includegraphics[scale=1]{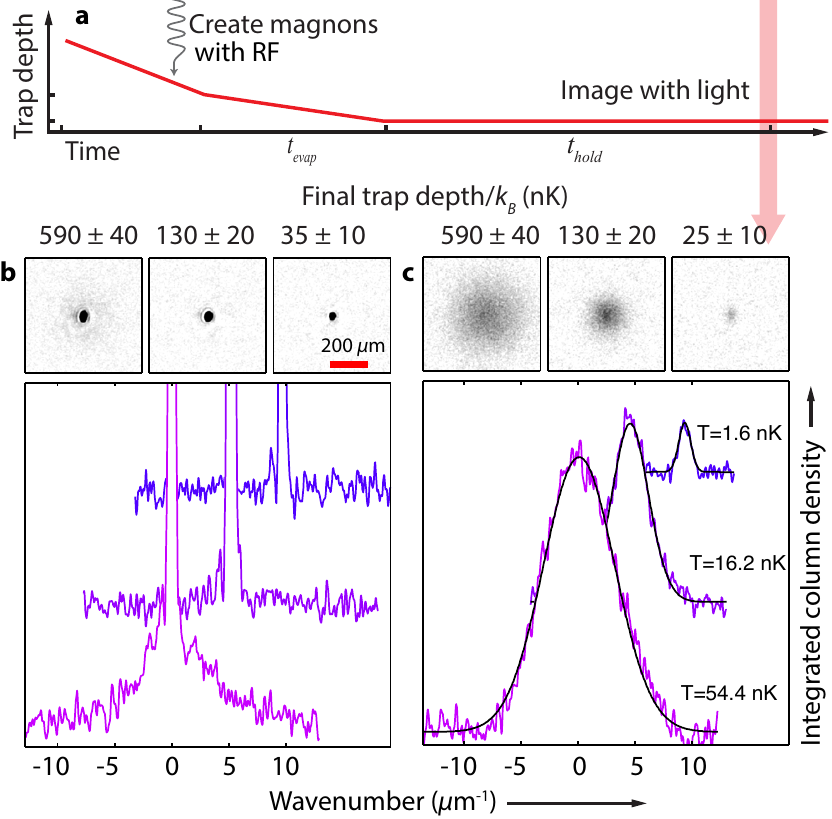}
\caption{\textbf{Magnon thermometry.} (a) Magnons are created and decohere rapidly in a non-degenerate spinor Bose gas at an intermediate trap depth. Forced evaporative cooling to a final, variable, trap depth reduces the temperature and the majority gas undergoes Bose-Einstein condensation. Images and corresponding integrated line profiles of the momentum distribution of the majority (b) and magnon gas (c) are each shown at three different final trap depths. Line profiles are shown offset for clarity. A condensate obscures the momentum distribution of the non-condensed fraction of the majority gas, especially at low temperatures. In contrast, the magnons can have have little to no condensed fraction, allowing non-condensed magnons to be identified and the temperature determined.} 
 \label{fig:scheme}
\end{figure}

Our procedure to measure temperatures is illustrated in Fig.~\ref{fig:scheme}(a). Experiments begin with spin-polarized \Rb in the $|F{=}1,m_F{=}{-}1\rangle$ state confined in an anisotropic optical dipole trap and cooled to just above quantum degeneracy by forced evaporation to an intermediate trap depth.  The spin quantization axis is defined by a 180 mG bias magnetic field. We tip the gas magnetization with a brief radio-frequency (RF) pulse, coherently transferring a small fraction (up to 15\%) of the atoms primarily into the $|F{=}1,m_F{=}0\rangle$ state, creating magnons which rapidly decohere and thermalize. The gas is then cooled further by forced evaporation to a final trap depth where the temperature reaches a steady state.

Next, we release the gas from the optical trap and image the momentum distribution of the magnons.  Upon extinguishing the trap light, the gas expands rapidly in the most tightly confined (vertical) direction, quickly reducing the outward pressure within the gas. We then use a sequence of microwave and optical pulses to drive away atoms not in the $|m_F{=}0\rangle$ Zeeman state.  Finally, we transfer the remaining atoms to an atomic state ($|F{=}2, m_F{=}1\rangle$) suitable to two-dimensional magnetic focusing \cite{tung10presuperfluid} so that their transverse spatial distribution, which we image, closely reflects their initial transverse velocity distribution. The majority gas is probed by imaging $|m_F{=}0\rangle$ atoms immediately after the application of an RF pulse.

\begin{figure*}[tb!]
\centerline{\includegraphics{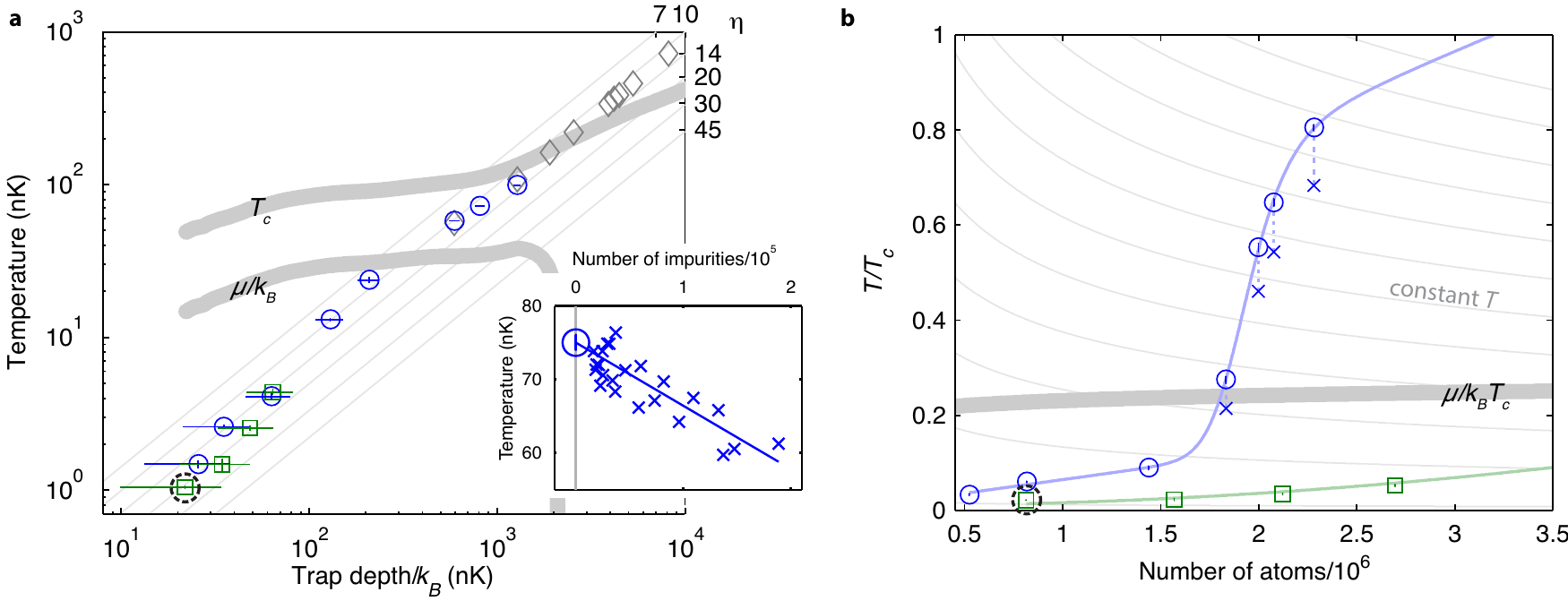}}
\caption{\textbf{Thermometry results.} Two runs (blue circles, green squares) differ in initial atom number. (a) $T$ and (b) $T/T_c$ are measured at various optical trap depths. Lower $T$ and $T/T_c$ are achieved in runs with larger initial atom number. In (a), thermometry using the majority spin population (light gray diamonds) agrees with thermalized magnon thermometry extrapolated to the zero-magnon values (circles and squares).  Error bars show statistical uncertainty of extrapolated temperature (vertical) and systematic uncertainty in trap depth (horizontal).  Thin diagonal gray lines show contours of $\eta$, the ratio of trapping potential depth to temperature. Thick gray lines show calculated $T_c$ and $\mu/k_B$ in (a) and $\mu/k_B T_c$, calculated using the frequencies of the $k_B\times60$ nK deep trap, in (b). (b) Efficiency of evaporation only (circles, squares), with error bars showing statistical uncertainty of the extrapolated $T/T_c$, is compared to magnon-assisted evaporation (``x''). Solid lines connecting points are guides to the eye, with a steeper slope indicating more efficient evaporation. Dotted lines connect corresponding points at the same trap depth.  Inset: the magnon-free temperature is determined by extrapolating single-shot temperatures (``x'') vs.\ total number of magnons imaged, here at trap depth of 800 nK, to the zero-magnon limit (circle). A similar extrapolation determines the value of $T/T_c$ achieved by evaporation only.}
\label{fig:TvsV}
\end{figure*}

The advantage of using incoherent spin excitations to measure temperature is exhibited in Fig.\ \ref{fig:scheme}, which compares the momentum distribution of the majority gas (b) to that of the magnons (c). At $T \lesssim 0.3 \, T_c$, the non-condensed fraction of the majority gas is obscured by the condensed fraction, the distribution of which is broadened owing to interactions and imaging resolution.  In contrast, the temperature can be easily extracted from the momentum distribution of the thermalized minority-spin gas.  Temperatures extracted from the magnons [Fig.~\ref{fig:TvsV}(a)] agree with those from the majority-spin distribution under conditions where both can be measured. At lower temperatures, the magnon thermometer allows measurements in a hitherto unmeasurable regime.

Our measurements reveal that forced evaporative cooling produces Bose gases with extremely low temperature and entropy.  At the lowest trap-depth setting (dashed circles, Fig.~\ref{fig:TvsV}), we measure an average gas temperature of $T=1.04(3)_\mathrm{stat}(7)_\mathrm{sys}$ nK, corresponding to $T/T_c = 0.022(1)(2)$ where $T_c$ is calculated using the measured atom number, $N = 8.1\times 10^5$, and optical trap frequencies.  For this gas, $k_B T / \mu \approx 0.07$, with $\mu$ the chemical potential, meaning that thermal excitations in the high density region of the condensed gas (in the majority spin state) are predominantly phonons. Measurement procedures, error estimates, and calculations of $T_c$ and $\mu$ are detailed in Methods.

We observe that evaporative cooling is highly efficient with respect to particle loss in the regime $\mu/k_B < T < T_c$ [Fig.~\ref{fig:TvsV}(b)]. When $T < \mu/k_B$, the cooling efficiency reduces, consistent with the fact that the chemical potential accounts for a non-negligible amount of the energy carried away by each atom lost to evaporation. The relevance of the chemical potential in evaporation is also indicated in Fig.~\ref{fig:TvsV}(a) by the observation that $\eta\equiv U/(k_BT)$, the ratio of trap depth to thermal energy, increases at lower trap depths, and that the temperature at a fixed trap depth depends on the number of atoms. In an evaporatively cooled gas, the temperature responds to the effective trap depth, $U_\mathrm{eff}=U-\mu$, the potential energy depth minus the chemical potential, rather than the trap depth alone, and in the regime $T <\mu/k_B$, the difference $U-U_\mathrm{eff}=\mu$ is significant.

We calculate the Bogoliubov energy spectrum of the confined quantum degenerate gas, including the effects of trap anharmonicity, and find its entropy at equilibrium (see Methods) to be $S/N = 1 \times 10^{-3}\, k_B$, the lowest value ever reported for an atomic gas.  By comparison, the 500 pK Bose gas reported in Ref.\ \cite{lean03pico} has $S/N = 3.6 \, k_B (T/T_c)^3 \sim 1.5 \, k_B$ using relations for a non-interacting gas at $T/T_c \sim 0.75$.  Other reported values include $S/N \sim 0.05 \, k_B$ at the center of a resonantly interacting Fermi gas \cite{ku12lambda}, $S/N = 0.27 \, k_B$\cite{bakr11blockade} or ${\sim}0.1 \, k_B$\cite{medl11demag} for bosons in a lattice, and $T/T_c\sim 0.15$ in a double-well Bose-Einstein condensate  \cite{Gati2006}.  Thermometry based on imaging incoherent phonons indicated a temperature around 7 times higher than reported here in a \Rb condensate of similar density \cite{Schley2013}.

Having magnons present during evaporative cooling reduces the temperature of the trapped gas by increasing the efficacy of evaporative cooling.  Forced evaporative cooling from a trap with an effective trap depth $U_\mathrm{eff} \gg k_B T$ has a cooling power proportional to the number of thermal excitations with excitation energies above $U_\mathrm{eff}$.  In a weakly interacting single-component degenerate Bose gas, the number of thermal excitations is determined by the temperature and is independent of the total particle number, fixing the evaporative cooling rate.  By seeding the gas with additional spin excitations at constant total particle number, the total number of thermal excitations, and thus the evaporative cooling power, increases.

We observe that, for $T \gtrsim \mu/k_B$, this magnon-assisted evaporation leads to $T/T_c$ lower than that reached by single-component evaporation at the same final particle number (``x'' points in Fig.\ \ref{fig:TvsV}(b)). When $T \lesssim \mu/k_B$, we do not find magnons to reduce the temperature, perhaps because of the small number of non-condensed magnons and the decrease in evaporative cooling efficiency in general. Magnon-assisted evaporation is also evidenced by the fact that the per-atom loss rate is higher for the magnon population than for the spin-majority population:  For both populations, the atom loss reflects the selective evaporative loss of thermal excitations.  For the spin-majority atoms, thermal excitations are just a small fraction of the total population, whereas they represent a much higher fraction of the magnons.

Unbiased thermometry of the majority gas is performed by extrapolating measurements with various numbers of magnons to the zero-magnon limit (Fig.~\ref{fig:TvsV}(a) inset).  In addition, by examining evaporative cooling in a state-dependent optical trap, we have confirmed that the temperature indicated by the magnons varies with the trap depth of the spin-majority gas, even for a fixed trap depth of atoms in the $|m_F{=}0\rangle$ state.  This observation gives further confidence that the magnon thermometer accurately records the temperature of the majority gas.

\begin{figure*}[tb!]
\centerline{\includegraphics{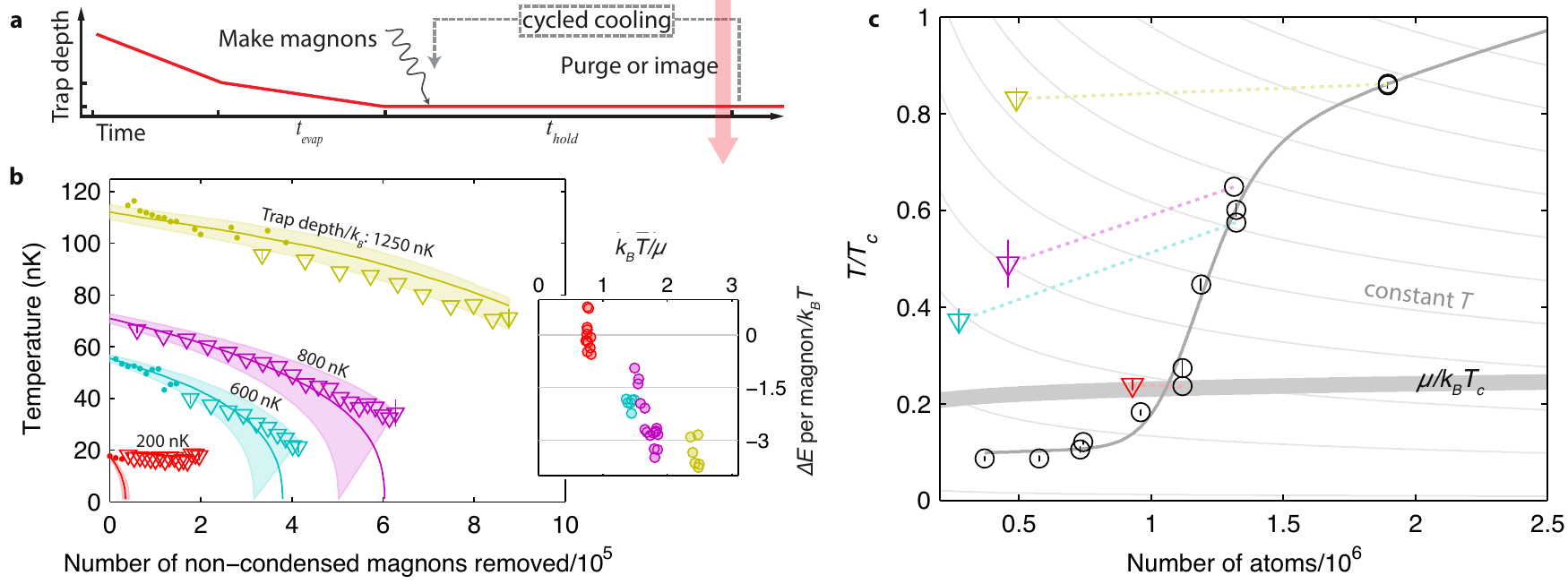}}
\caption{\textbf{Cycled decoherence cooling.} (a) Magnons are created at the final trap depth and cool the gas as they thermalize. Thermalized magnons can be removed from the trap and the cooling process repeated before measuring the temperature by imaging the momentum distribution of the thermalized magnons.  (b) Each non-condensed magnon removed from the trap takes energy from the gas. Decoherence cooling trajectories are plotted versus cumulative number of magnons removed. Closed circles show temperatures after a single cycle of decoherence, but with varying numbers of magnons. Some additional magnon-assisted evaporative cooling may also be present. Open triangles show repeated cycles of magnon creation, thermalization, and purge, with representative error bars on first and last triangles giving statistical error over several repetitions. Solid lines show zero-free-parameter theory predictions assuming each non-condensed magnon removes 3 $k_BT$ energy. Solid patches indicate the range of predictions included within the uncertainty in the cumulative number of magnons removed (we image $75\% \pm 10\%$ of the magnons present). Inset: the amount of energy removed by each magnon, in units of $k_B T$, is plotted versus the ratio of $\mu/k_B T$, where each point is based on an estimate of the slope over 4 consecutive triangle points. We observe that the net energy carried away by impurities vanishes when $T \lesssim \mu/k_B$. (c) Efficiency of evaporation only (black circles) is compared to cycled decoherence cooling (open triangles), with error bars representing statistical uncertainty. The solid line between circles is a guide to the eye. Dotted lines connect corresponding points at the same trap depth. Compared to evaporation, decoherence cooling can reach considerably lower $T/T_c$ at the same trap depth in the regime $T\gtrsim\mu/k_BT$ and $T/T_c\lesssim 0.9$.}
\label{fig:magnoncooling}
\end{figure*}
 
The same characteristics that make dilute magnons a good thermometer---high relative energy and entropy per particle---enable them to cool without lowering the trap depth, beyond what can be achieved by magnon-assisted evaporation. Spin excitations created in a degenerate Bose gas decrease its temperature by a process known as decoherence \cite{lewa03cool} or demagnetization \cite{Fattori2006,Medley2011} cooling. Immediately after the RF pulse is applied, the magnon population has the same energy and momentum distribution as the initially polarized degenerate Bose gas, with a large condensed fraction, which carries no energy above the chemical potential, and a small normal fraction, which carries on the order of $k_B T$ of energy per particle. Upon thermalization, the normal fraction within the magnon gas increases, bounded from above by its critical number for condensation, and the energy and entropy of the magnons increases. The energy gained by the magnons is energy lost by the majority gas, as the entire process, apart from the slight effects of magnetic field inhomogeneities and dipolar interactions, occurs at constant energy \footnote{In Fermi gas systems, a similar process leads to heating as the energy and entropy per particle of the minority spins is less than that of the majority gas at equilibrium.}.

While the overall entropy rises during thermalization, the entropy per particle of the majority gas may decrease. Considering a non-interacting harmonically trapped gas, we calculate that decoherence cooling would reduce $T/T_c$ of the majority gas only when $T/T_c  < \sqrt[3]{(3 \zeta(3))/(4 \zeta(4))} = 0.94$, consistent with our observations.

To demonstrate the capabilities of decoherence cooling, we introduce a variable number magnons at the final trap depth and allow them to evolve toward thermal equilibrium with the assistance of a weak magnetic field gradient of 0.2--2 mG/cm (for more on magnon thermalization, see Supplemental Discussion). Then, as illustrated in Fig.~\ref{fig:magnoncooling}(a), we either measure the temperature or expel the magnons using a spin-selective process, leaving a colder spin-polarized condensed gas.  The decoherence cooling process can then be repeated to further cool the sample, with the number of magnons, magnetic field gradient, and thermalization time optimized at each repetition. A single pass of decoherence cooling, using pseudo-spins in a degenerate \Rb gas, has previously been shown to slightly reduce the temperature, but not $T/T_c$, of the resulting mixture \cite{lewa03cool}.

Whereas the thermal energy $k_B T$ achieved by forced evaporative cooling of a single-component gas is limited to around $1/\eta \sim 1/10$ of the trapping potential depth, the temperatures reached by cycled decoherence cooling are far lower. For example, we produced a 20 nK gas within a $k_B\times600\,\mathrm{nK}$ deep potential, corresponding to $1/\eta = 1/30$, and reduced $T/T_c$ substantially, from 0.58 to 0.37 [cyan triangles, Fig.~\ref{fig:magnoncooling}(b) and (c)]. On the other hand, decoherence cooling is generally less efficient in terms of atom loss.  An atom lost through evaporative cooling removes an energy $\sim\eta k_B T$, whereas a magnon brought to equilibrium and forcibly ejected only removes 3 $k_B T$ for a non-interacting gas, consistent with our observations when $k_B T \gg \mu$. At lower temperatures, we expect each impurity to remove less energy, 1 to $1.5\,k_BT$, as interactions and quantization of the trap modes become more important. Even so, cooling would be expected, in the ideal case, to become increasingly effective as the heat capacity drops. In contrast, we find decoherence cooling to be ineffective at reducing both $T$ and $T/T_c$ in the low temperature regime $k_BT\lesssim \mu$ (Fig.~\ref{fig:magnoncooling}(b, c), inset and red triangles), perhaps owing to energy deposited in the trapped gas during the magnon expulsion and by heating processes, which become more significant as thermalization times increase.

In principle, the process of creating the spin mixture employed in decoherence cooling can heat the gas by changing its interaction energy. However, neglecting small magnetic dipolar interactions, the magnon is a gapless Goldstone excitation corresponding to the rotational symmetry of real spins  \cite{Marti2014}. Accordingly, the $s$-wave scattering lengths that characterize the interactions between two majority-spin atoms, $a_{{-}1,{-}1}$, and between a majority and minority-spin atom, $a_{{-}1,0}$, are identical. In contrast, spin excitations involving a different hyperfine state can deposit energy into the condensate due to a mismatch of $s$-wave scattering lengths.

In this work, we have demonstrated the use of spin excitations within a highly degenerate Bose gas to measure and reduce the temperature.  With our improved thermometry, we observe a record low entropy of $1\times 10^{-3}\, k_B$ per particle for a spin-polarized Bose-Einstein condensate, illustrating the power of standard cooling techniques to reach the long-sought low entropy regime. Magnon-assisted evaporation and cycled decoherence cooling should work for other spin or pseudospin mixtures of Bose gases so long as the gas thermalizes at constant minority particle number, providing a flexible platform for creating or maintaining low entropies in deep traps.

\section*{Acknowledgments}
We thank Holger Kadau and Eric Copenhaver for assistance improving the experimental apparatus. We acknowledge the primary research support from NASA and the AFOSR through the MURI program, and also secondary support for personnel through the NSF.  G.E.M. acknowledges support from the Fannie and John Hertz Foundation.

\putbib[magnoncool,allrefs_x2,magnoncoolNotes]
\end{bibunit}

\section*{Author contributions}
All authors provided experimental support and commented on the manuscript. Experimental data were acquired by R.O.\ and F.F.\ and analyzed by R.O. G.E.M.\ conceived and performed preliminary experiments with cycled decoherence cooling. The manuscript was prepared by R.O.\ and D.M.S-K. R.O.\ performed the calculations of entropy per particle. D.M.S-K.\ supervised all work.

\section*{Competing financial interests}
The authors declare no competing financial interests.

\begin{bibunit}[naturemagnourl]

\renewcommand{\theequation}{M\arabic{equation}}
\renewcommand{\thefigure}{M\arabic{figure}}
\renewcommand{\bibnumfmt}[1]{[M#1]}
\renewcommand{\citenumfont}[1]{M#1}

\section*{Methods}

\subsection{Optical trapping}

Experiments were performed on \Rb gases in a single-beam optical dipole trap  \cite{Marti2014}.  Evaporative cooling was realized by gradually lowering the depth of the optical trap.  The optical trap was formed by light with a wavelength of 1064 nm, brought to a cylindrical focus at the location of the atoms.  The optical trap was highly anisotropic, with trapping frequencies having a typical ratio 1:15:140, with the tightest confinement in the vertical direction.  At low optical powers, the optical dipole trap is influenced by gravity so that the ratio of trap frequencies differs somewhat.  Trap frequencies were measured empirically at several optical powers, extrapolated between measurements, and confirmed by comparing with a simple model of the trapping potential that accounts for the Gaussian focus and force of gravity.

The bulk of our experiments were performed with linearly polarized trap light, which imparts equal dipole force on all Zeeman sublevels of $F{=}1$ \Rb. To verify that our temperatures vary with the trap depth of the majority gas, rather than that of the minority spins, we also performed experiments with circularly polarized trap light, which applies a differential force to atoms in different Zeeman sublevels. Using this circularly polarized trap, we vary the trap depth of the majority gas without affecting that of the $|m_F{=}0\rangle$ atoms by preparing the majority gas in either of the $|m_F{=}{+1}\rangle$ or $|m_F{=}{-1}\rangle$ states.

The lifetime and heating rate of the atoms in the optical trap are consistent with spontaneous scattering of the far-detuned trapping light.

\subsection{State purification and preparation}\label{sec:stateprep}

About one millisecond after extinguishing the optical trapping light, we initiate the imaging sequence by purging atoms in the $|F{=}1, m_F{=}{-1}\rangle$ state---the majority component of our gas in most of our experiments--- using a series of alternating microwave pulses (resonant with the $|F{=}1, m_F{=}{-1}\rangle \rightarrow |F{=}2, m_F{=}0\rangle$ transition) and light pulses (resonant with the $F{=}2$ hyperfine manifold).

The number and strength of the optical/microwave pulses is optimized to reliably expel all of the $|F{=}1, m_F{=}{-1}\rangle$ atoms with minimal impact on the number of atoms in the $|F{=}1, m_F{=}0\rangle$ state. Regardless, the purge process removed between 10\% and 40\% of the  $|F{=}1, m_F{=}0\rangle$ atoms, depending on the number of $|F{=}1, m_F{=}{-1}\rangle$ atoms being purged and other experimental parameters, but it does not visibly impact the momentum of the atoms that remain. We estimate the number of magnons from the images by accounting for this loss, allowing considerable uncertainty.

Finally, we employ a calibrated microwave sweep to transfer atoms in the $|F{=}1, m_F{=}0\rangle$ state to the magnetically trappable $|F{=}2, m_F{=}1\rangle$ state with greater than 95\% efficiency.

\subsection{Magnetic focusing}\label{sec:magfocus}

The magnetic focusing lens takes the form of a magnetic potential with negligible curvature along the vertical (imaging) axis and weak, harmonic curvature in the image plane. The resulting magnetic trap causes atoms with momenta $\vec{p}$ in the image plane to converge on the real-space point $\vec{x}_p=\vec{p}/m \omega$ after a time $t_\mathrm{foc}=2\pi/4 \omega$, where $m$ is the atomic mass and $\omega$ the trap frequency for atoms in the $|F{=}2, m_F{=}1\rangle$ state. The vertical gradient is selected to cancel the effect of gravity, and the trap frequency $\omega$ is chosen such that the thermal atomic momentum-space cloud is resolvable by our imaging system at the lowest accessible temperatures while still allowing us to image temperatures near $T_c$.

Uncertainty in the parameter $\omega$ is the largest source of systematic error in the temperatures reported in this work. Our measurements found $\omega = 2\pi\times 2.88 \pm 0.09$~rad/s and the relation  $\vec{x}_p=\vec{p}/m \omega$ was verified by comparing the amplitude and phase of center-of-mass oscillations of our gas imaged both \textit{in situ} and after application of the magnetic focusing lens.

Time-of-flight expansion of the gas in the unconfined vertical direction along with misalignment of the trap, imaging, and focusing axes can lead to several types of aberration. For example, any projection of the vertical extent of the gas into the imaging plane will manifest as additional apparent momenta in the imaged column density. Such aberrations can lead to an overestimate of the temperature of the gas and are most pertinent at low temperatures. Trap, imaging, and focusing axes are all aligned to gravity in our system, and can be verified by imaging highly degenerate gasses, which expand negligibly in the direction of weakest optical confinement.

\subsection{The imaging system}\label{sec:imaging}

After magnetic focusing, the atoms are imaged with a 40 to 100 $\mu$s pulse of light resonant with the $F{=}2 \rightarrow F{=}3$ cycling transition.  The magnification of our imaging system was calibrated using an optical micrometer target that was placed in a plane equivalent to the plane of the atoms, with respect to the imaging system. This calibration was repeated several times with fractional uncertainty of 1\% and is included in the systematic error of our thermometer.

The resolution of our imaging system, approximately 8 $\mu$m, corresponds to a temperature of about 0.2 nK. We do not correct extracted temperatures for finite imaging resolution.

\subsection{Details of temperature fitting procedure}\label{sec:temp}

Atom column momentum densities $n_p$ are fit to a Bose-enhanced momentum distribution $n_p = b + A g_j(e^z)$, with $g_j$ the Bose function (polylogarithm) of order $j$ and the argument
\begin{equation*}
z = \alpha - (p_w-p_{w0})^2/k_B T_w-(p_t-p_{t0})^2/k_B T_t.
\end{equation*}
For non-degenerate gasses, $\alpha = \mu/k_B T$ is a free parameter of the fit, along with the zero of momentum $(p_{w0}, p_{t0})$, the background level $b$, peak level $A$, and temperatures $T_w$ and $T_t$ along the weak and tight in-plane axes of the optical trap, respectively. For degenerate gasses, $\alpha = 0$.

Generally, $T_w = T_t$ within the error of the fit, however at low temperatures, several effects cause $T_t$ be an unreliable estimate of the gas temperature. Although the condensate expands primarily along the unfocused vertical direction, at the very low temperatures reported in this work, condensate expansion along the more tightly confined in-plane direction can be manifest. Also, the tilt of the trap in the tight direction is more difficult to calibrate, leading to a (fictitious) systematic upward shift in the apparent $T_t$ by the mechanism explained in Sec.~\ref{sec:magfocus}. Finally, at the lowest temperatures reported in this work, the semiclassical condition $k_B T \gg \hbar \omega$ does not hold along the tight axis as it does along the weak axis ($\hbar \{\omega_w,\omega_t\}/k_B \sim \{0.05, 1\}\, \mathrm{nK}$). Thus, in this work, we estimate temperatures by $T=T_w$ alone.

All fits exclude regions of the column density that include condensed atoms, and each fit was performed many times, varying both the size of the excluded central region and the order $j\in[0.5,2]$ of the Bose function in order to look for systematic shifts in the temperature. The high-momentum tails of the momentum distribution $n_p$ are insensitive to the particular Bose function employed ($g_j(e^z) \rightarrow e^z$ as $e^z\rightarrow 0$). We used only fits where the variation of the temperature with respect to the size of the exclusion region and the order of the Bose function was negligible compared to other sources of systematic error.

The uncertainty with which individual fits estimate the temperature is consistent with fundamental sources of noise, photon and atom shot noise, and the short-time (same day) shot-to-shot variation of estimated temperatures was consistent with the single-shot noise-limited error estimated by the fitting routine. This justifies the use of an uncertainty-weighted average of many independent measurements to compute the precise low temperatures reported in this work.

\subsection{Calculation of the entropy per particle}\label{sec:calc-entr-per}

Calculations of the entropy per particle cannot rely on the local density approximation because the thermal phonon wavelength is far larger than the transverse extent of the condensed gas.  We can, however, apply a one-dimensional local density approximation along the long axis of the condensate.

To calculate the entropy, we numerically solve for the Gross-Pitaevskii ground state $\Psi$ and Bogoliubov spectrum $\{\epsilon_i\}$ of small amplitude excitations in a weakly-interacting \Rb Bose gas with chemical potential $\mu$ in a trap with confinement in 2 dimensions, including the effects of gravity and anharmonicity \footnote{Neglecting gravity and approximating the trap as harmonic results in underestimating the entropy per particle by a factor of 1.6.}, and no confinement in the 3rd dimension.  To do this calculation, we discretize the 2-dimensional Gross-Pitaevskii equation and Bogoliubov equations on a Lagrange mesh based on the Hermite polynomials \cite{Baye1986,McPeake2002a}.  The Bogoliubov equations include the substitution $\mu \rightarrow \mu-\hbar^2 k^2/2m$, for relevant values of the wave vector $k$, to account for excitations in the unconfined longitudinal direction. We verify that our calculation of the excitations $\{\epsilon_i\}$ are accurate by comparing to analytic results in regimes where they are available, for example in the harmonic non-interacting limit and in the Thomas-Fermi limit with cylindrical symmetry.

The entropy and number of atoms per unit length can then be calculated using the equilibrium relations
\begin{eqnarray*}
\frac{S}{L} &=&\sum_i \frac{\left(\epsilon_i-\mu\right)/k_B T}{e^{\left(\epsilon_i-\mu\right)/k_B T} - 1} - \log\left(1 - e^{\left(\mu-\epsilon_i\right)/k_B T}\right)\\
\frac{N}{L} &=& N_c + \sum_i \frac{1}{e^{\left(\epsilon_i-\mu\right)/k_B T} - 1},
\end{eqnarray*}
with $N_c = \Psi^\ast\Psi$.
By fixing $T$ and varying $\mu$ ($\Psi$ and $\{\epsilon_i\}$ are recalculated for each value of $\mu$), we can consider the entropy and number per unit length as functions of $\mu$: $\frac{S}{L}(\mu)$ and $\frac{N}{L}(\mu)$.

We use the local density approximation and Thomas-Fermi approximation along the long axis of our trap (with trap frequency $\omega_z$) to write the total entropy of the gas as
\begin{equation*}
S = \int_{-\infty}^{\infty} \frac{S}{L}\left(\mu_\mathrm{max} - \frac{1}{2} \omega_z^2 z^2\right) dz,
\end{equation*}
and the total number of atoms as
\begin{equation*}
N = \int_{-\infty}^{\infty} \frac{N}{L}\left(\mu_\mathrm{max} - \frac{1}{2} \omega_z^2 z^2\right)  dz,
\end{equation*}
where $\mu_\mathrm{max}$ is chosen to yield the experimentally measured value of $N$. The entropy per particle of the gas is then $S/N$.

At the lowest trap depths, the rate at which the magnons are lost by forced evaporation becomes comparable to the rate at which non-condensed magnons are generated through thermalization. We speculate that the number of thermal excitations in the spin-majority gas under these conditions is similarly lower than at the observed temperature in equilibrium.  As such, our estimate of $S/N = 1\times 10^{-3} k_B$ may be an underestimate of the true steady-state entropy.

\subsection{Calculating $T_c$ and $\mu$}\label{sec:calculating-t_c-mu}

The estimates of $T_c$ and $\mu$ reported use standard formulas for Bose gases in the non-interacting and Thomas-Fermi limits, respectively \cite{Pitaevskii2003}. The effects of trap anharmonicity and finite interactions on both $\mu$ and $T_c$ were considered and are on the order of the systematic uncertainty in our temperature measurements.

\subsection{Theoretical limits for decoherence cooling} \label{sec:limits-decoh-cool}

Considering an ideal Bose gas in a harmonic trap with geometric-mean frequency $\omega$, the following relations for the saturated thermal atom number $N_\mathrm{th}$, energy $E$, and heat capacity $C$, at the temperature $T$, apply:
\begin{subequations}
\begin{align}
N_\mathrm{th} &= \zeta(3)\left(\frac{k_BT}{\hbar\omega}\right)^3\\
E &= 3 \zeta(4)\frac{(k_BT)^4}{(\hbar \omega)^3} = \frac{3 \zeta(4)}{\zeta(3)}k_B T N_\mathrm{th}\\
C &= \frac{dE}{dT}=12 k_B \frac{\zeta(4)}{\zeta(3)}N_\mathrm{th}.
\end{align}
\label{eq:ideal-relations}
\end{subequations}
Suppose we transfer $dN>0$ atoms to the minority spin state (where $dN$ is far below the critical number for magnon condensation), let them thermalize, and eject them. To lowest order, the majority gas loses energy $dE=-3k_BTdN$, and thus the temperature changes by $dT=dE/C=-3(k_BT)/C\,dN$. The non-condensed fraction changes as, using Eqs.~\ref{eq:ideal-relations},
\begin{subequations}
\begin{align}
d\left(\frac{N_\mathrm{th}}{N}\right) &= \frac{1}{N}\left[dN_\mathrm{th}+\frac{N_\mathrm{th}}{N}dN\right]\\
&= \frac{dN}{N}\left[-\frac{3\zeta(3)}{4\zeta(4)}+\frac{N_\mathrm{th}}{N}\right].
\end{align}
\end{subequations}
Thus, we must have $N_\mathrm{th}/N<\left(3\zeta(3)\right)/\left(4\zeta(4)\right)=0.83$ to reduce the non-condensed fraction through such cooling. This corresponds to, for the case under consideration, $T/T_c<(0.83)^{1/3}=0.94$.


\putbib[magnoncool,allrefs_x2,magnoncoolNotes]
\end{bibunit}

\end{document}


\title{Supplementary material: Thermometry and cooling of a Bose-Einstein condensate to 0.02 times the critical temperature}
\author{Ryan Olf}
\email{ryan@efrus.com}
\affiliation{Department of Physics, University of California, Berkeley, California 94720, USA}
\author{Fang Fang}
\affiliation{Department of Physics, University of California, Berkeley, California 94720, USA}
\author{G. Edward Marti}
\altaffiliation[Present address: ]{JILA, National Institute of Standards and Technology and University of Colorado, Boulder, Colorado 80309-0440, USA; edward.marti@jila.colorado.edu}
\affiliation{Department of Physics, University of California, Berkeley, California 94720, USA}
\author{Andrew MacRae}
\affiliation{Department of Physics, University of California, Berkeley, California 94720, USA}
\author{Dan M. Stamper-Kurn}
\affiliation{Department of Physics, University of California, Berkeley, California 94720, USA}
\affiliation{Materials Sciences Division, Lawrence Berkeley National Laboratory, Berkeley, California 94720, USA}
\date{May 21, 2015} 

\maketitle

\section*{Discussion}
\subsection*{Magnon thermalization}\label{sec:magn-therm}

\begin{figure*}[tb!]
  \centering
  \includegraphics{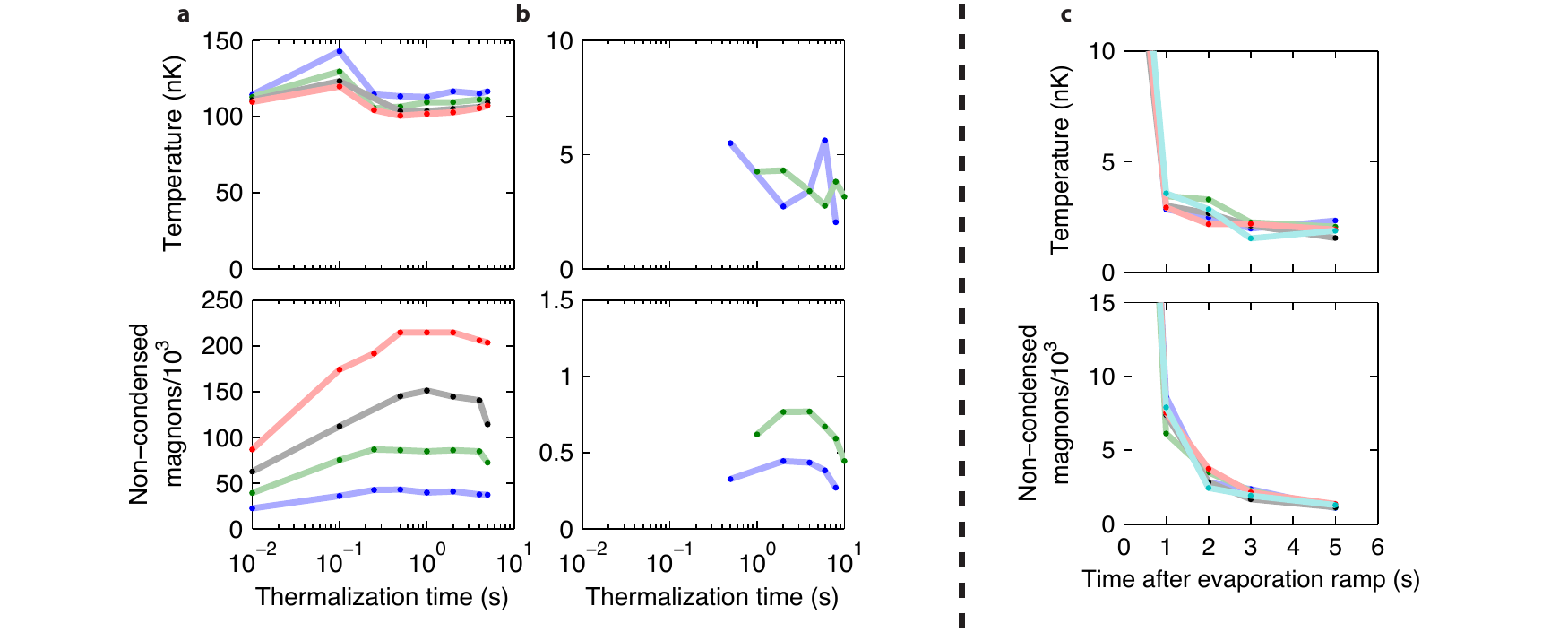}
  \caption{\textbf{Magnon thermalization.} Magnons created in a degenerate gas with (a) $k_BT\sim 2\mu$  and (b) $k_BT\sim\mu/3$ thermalize at trap depths of approximately $k_B\times \text{1,200}$ and 35 nK, respectively, in the presence of a weak magnetic field gradient. Throughout, traces (colored points and connecting lines) differ in initial number of magnons. In (a), the momentum distribution of the non-condensed magnons can be determined as soon as they are created. The indicated temperature reaches a steady state and the number of non-condensed magnons saturates after a few hundred milliseconds. The steady-state temperature depends on the number of magnons created owing to the effects of decoherence-driven cooling.  In (b), temperatures and the number of non-condensed magnons cannot be determined until after nearly one second of thermalization, at which time the temperature has already reached steady state. (c) Magnons created before the final forced evaporation ramp are incoherent throughout the whole ramp and reach a steady-state temperature $k_BT\sim\mu/3$ after two to three seconds at the final trap depth of approximately $k_B\times 25\,\mathrm{nK}$. }
  \label{fig:magn-therm}
\end{figure*}

We found that applying a magnetic field gradient was necessary to achieve thermalization between initially coherent, quantum degenerate spin components, but can be chosen to be small enough to impart negligible energy. A gradient of 0.2 mG/cm, typical in our experiments on decoherence cooling, corresponds to an energy of approximately $h\times 10\,\mathrm{Hz} = k_B \times 0.5$ nK across the longest dimension of our gas and was sufficient to produce a saturated non-condensed fraction of the $|m_F{=}0\rangle$ state in 100--2000 ms, depending on the temperature of the gas. Thermalization of magnons created out of a degenerate gas in two representative cases of $k_BT/\mu < 1$ and $k_BT/\mu >1$ are shown in Fig.~\ref{fig:magn-therm} (a) and (b). Fig.~\ref{fig:magn-therm} (c) shows the extracted temperature of magnons created at a higher trap depth (in a non-degenerate gas) reaching a steady state following the final phase of evaporative cooling, for a representative case $k_BT/\mu < 1$. Temperatures extracted from magnons created from a non-degenerate gas before completing evaporation and from magnons that are created in the degenerate gas after evaporation agree with each other, when the same experimental parameters are employed, and when extrapolated to the zero-magnon limit.

The physics behind the decoherence and thermalization of spin excitations, and the variables that affect it, remains a compelling direction of future research. Previous experiments on degenerate $F{=}1$ \Rb gases have found magnetic field inhomogeneity to facilitate decoherence and thermalization of spin populations, typically on timescales of several 10s--100s of milliseconds \cite{Chang2005,Lewandowski2003,Kronjager2005}. We have previously observed coherence between magnons in different momentum states out to several hundred milliseconds \cite{Marti2014} with a negligible gradient. However, none of these experiments claim to have operated in the highly-degenerate phonon-dominated regime $k_BT<\mu$. With very small gradients, we find the zero-momentum magnons created in this work can remain coherent for many seconds. In strongly-interacting Fermi gases, decoherence of spin excitations has been well characterized and is limited by a fundamental lower-bound on the spin diffusivity, $\hbar/m$ \cite{Bardon2014,Koschorreck2013}. In this context, a gradient is essential to drive the diffusive spin transport. 

When $k_BT<\mu$, the thermal excitations of a homogeneous Bose-Einstein condensate are phonon-like, and thus thermalization of the free-particle-like magnons with the majority gas requires collisions between quasi-particles with very different dispersion. Such interactions are expected in the first order beyond the Bogoliubov mean-field theory (the Beliaev theory) \cite{Phuc2013}, though to our knowledge no one has produced an estimate or direct measurement of the thermal phonon-magnon cross section. We speculate that thermalization of the magnons occurs primarily at the edges of the condensate where low condensate density implies a better match between majority spin and minority spin dispersion. In addition to facilitating decoherence, the applied gradient may aid in thermalization by driving the transport of low-energy spin excitations to the edge of the condensate.

\bibliographystyle{naturemagnourl}
\bibliography{magnoncool}